\documentclass[showpacs,aps,prd]{revtex4}
\input epsf

\begin{document}

\title{New molecular candidates: $X(1910)$, $X(2200)$, and $X(2350)$}
\author{Jian-Rong Zhang}
\author{Guo-Feng Chen}
\affiliation{Department of Physics, National
University of Defense Technology, Hunan 410073, China}

\date{\today}

\begin{abstract}
Assuming the newly observed resonant structures $X(1910)$, $X(2200)$, and $X(2350)$ as
$\omega\omega$, $\omega\phi$, and $\phi\phi$ molecular states respectively, we
compute their mass values in the framework of QCD sum rules.
The numerical results are $1.97\pm0.17~\mbox{GeV}$ for $\omega\omega$ state, $2.07\pm0.21~\mbox{GeV}$
for $\omega\phi$ state, and
$2.18\pm0.29~\mbox{GeV}$
for $\phi\phi$ state,
which coincide with the experimental values of $X(1910)$, $X(2200)$, and $X(2350)$,
respectively. This supports the statement that
$X(1910)$, $X(2200)$, and $X(2350)$ could be
$\omega\omega$, $\omega\phi$, and $\phi\phi$ molecular candidates respectively.
\end{abstract}
\pacs {11.55.Hx, 12.38.Lg, 12.39.Mk}\maketitle

\section{Introduction}\label{sec1}
Very recently, Belle Collaboration reported
observations of new resonant structures at
$M(\omega\omega)\sim1.91~\mbox{GeV}$,
$M(\omega\phi)\sim2.2~\mbox{GeV}$,
and $M(\phi\phi)\sim2.35~\mbox{GeV}$
in $\gamma\gamma\rightarrow X\rightarrow\omega\omega$,
$\omega\phi$,
and $\phi\phi$ respectively \cite{X}.
For convenience, these resonances are called $X(1910)$, $X(2200)$, and $X(2350)$ here.
Since they
all decay into two light vector mesons, it is natural to suppose
them to be molecular bound states composed of two light mesons.
In theory, the molecular concept was put forward long ago
\cite{Voloshin} and it was predicted that
molecular states have a rich spectroscopy in Ref. \cite{Glashow}.
The possible deuteron-like two-meson bound
states were studied in Ref. \cite{NAT}.
Although molecular states have not been
confirmed in experiment, there already have had some candidates for
them.
For instance,
$Y(4260)$ could be
a $\chi_{c}\rho^{0}$ \cite{Y4260-Liu} or
an $\omega\chi_{c1}$  state \cite{Y4260-Yuan};
$Z^{+}(4430)$ could be a $D^{*}\bar{D}_{1}$ state
\cite{theory-Z4430n,theory-Z4430};
$Y(3930)$ is proposed to be a $D^{*}\bar{D}^{*}$
 \cite{theory-Y3930,Liu,3930-Ping}; $Y(4140)$ is interpreted as
a $D_{s}^{*}\bar{D}_{s}^{*}$ \cite{Liu,theory-Y4140};
$X(4350)$ could be a $D_{s}^{*}D_{s}^{0*}$
\cite{X4350-Zhang,X4350-Ma};
$Y(4274)$ could be a $D_{s}D_{s0}(2317)$ \cite{Y4274-Liu}.
For more molecular candidates,
one can also see some other Refs., e.g. \cite{reviews,reviews1}.
If molecular states can be completely confirmed by experiment, QCD
will be further testified and then one will comprehend the QCD low-energy
behaviors more deeply.
Therefore, it is interesting to
study whether the newly observed $X$ states could be molecular states.

In the real world,
quarks are confined inside hadrons and the strong
interaction dynamics of hadronic systems
is governed by nonperturbative QCD effect completely.
Many questions concerning dynamics of the quarks and gluons at large distances remain
unanswered or, at most, understood only at a qualitative level.
It is a great challenge to extract hadronic information quantitatively
from the rather simple Lagrangian of
QCD.
Fortunately, one can apply the QCD sum
rule method \cite{svzsum} (for reviews
see \cite{overview1,overview2,overview3,overview4} and references
therein), which is a nonperturbative
formulation firmly based on QCD basic theory and has been successfully employed to some light four-quark states \cite{HXChen,HXChen1,HXChen2,Zs,ALZhang,ZGWang,Nielsen}.
In Ref. \cite{qqqq}, the authors have studied the tetraquark state $qq\bar{q}\bar{q}$ by constructing and analyzing the sum rule composed of a diquark-antidiquark current with the quantum number $J^{P}=0^{+}$ and found masses of the tetraquark state $qq\bar{q}\bar{q}$
appear in the region of $0.6\sim1~\mbox{GeV}$, which are much lower than the mass of $X(1910)$.
Thereby, it may not likely to be a $qq\bar{q}\bar{q}$ tetraquark state for $X(1910)$.
In Ref. \cite{qqss},
the authors have studied the tetraquark $ud\bar{s}\bar{s}$ of $J^{P}=0^{+}$ in the QCD sum rule and the mass of the tetraquark turns
out to be around $1.5~\mbox{GeV}$, which is much lower than the mass of $X(2200)$.
Thus, it may not likely to be a tetraquark $ud\bar{s}\bar{s}$ for $X(2200)$.
In Ref. \cite{ssss}, the authors
have predicted the mass of $ss\bar{s}\bar{s}$ tetraquark state of $J^{P}=0^{+}$
to be about $2.2~\mbox{GeV}$ in the relativistic quark model,
which is slightly lower than the mass of $X(2350)$.
Just from the slight mass difference,
one may not judge that $X(2350)$ is unlikely to be a $ss\bar{s}\bar{s}$ tetraquark state.
However, one could at least see that the result does not exclude other
possible interpretations such as molecular picture for $X(2350)$.
Therefore, we intend to obtain mass information of
$\omega\omega$, $\omega\phi$, and $\phi\phi$
bound states from QCD sum rules,
and
investigate whether $X(1910)$, $X(2200)$, and $X(2350)$ could be
new molecular candidates.

The rest of the paper is organized as three parts. We discuss QCD
sum rules for molecular states in Sec. \ref{sec2}, with the similar procedure
as our previous works \cite{Zhang}.
The numerical analysis is made in Sec. \ref{sec3},
and masses of $\omega\omega$, $\omega\phi$, and $\phi\phi$ states are extracted
out. The Sec. \ref{sec4} includes
a brief summary and outlook.

\section{QCD sum rules for $\omega\omega$, $\omega\phi$, and $\phi\phi$ molecular states}\label{sec2}
The starting point of the QCD sum rule is to construct
the interpolating current properly and then write down the correlator. In full QCD,
the interpolating current for light vector meson can be found
e.g. in Ref. \cite{reinders}.
One can construct the molecular state current from
meson-meson type of fields. Meanwhile, note that
Belle Collaboration have indicated that
there are substantial $0^{+}$
components in all three modes ($\gamma\gamma\rightarrow X\rightarrow\omega\omega$,
$\omega\phi$,
and $\phi\phi$).
Thus, following forms of currents with $J^{P}=0^{+}$ are
constructed for $\omega\omega$, $\omega\phi$, and $\phi\phi$
\begin{eqnarray}
j_{\omega\omega}&=&(\bar{q}_{c}\gamma^{\mu}q_{c})(\bar{q}_{c'}\gamma_{\mu}q_{c'}),
\end{eqnarray}
\begin{eqnarray}
j_{\omega\phi}&=&(\bar{q}_{c}\gamma^{\mu}q_{c})(\bar{s}_{c'}\gamma_{\mu}s_{c'}),
\end{eqnarray}
\begin{eqnarray}
j_{\phi\phi}&=&(\bar{s}_{c}\gamma^{\mu}s_{c})(\bar{s}_{c'}\gamma_{\mu}s_{c'}),
\end{eqnarray}
where $q$ denotes light quarks $u$ and $d$,
with $c$ and $c'$ are color indices.
One should note that meson molecules in the real world are long objects
in which the quark pairs are far away from each other.
The currents in this work and in most of the QCD sum rule works
are local and the four field operators here
act at the same space-time point. It is a limitation inherent
in the QCD sum rule disposal of the hadrons since the bound states are
not point particles in a rigorous manner.
The two-point correlator is defined as
$\Pi(q^{2})=i\int
d^{4}x\mbox{e}^{iq.x}\langle0|T[j(x)j^{+}(0)]|0\rangle$.
In phenomenology,
the correlator
can be expressed as
\begin{eqnarray}
\Pi(q^{2})=\frac{\lambda_{H}^{2}}{M_{H}^{2}-q^{2}}+\frac{1}{\pi}\int_{s_{0}}
^{\infty}ds\frac{\mbox{Im}\Pi^{\mbox{phen}}(s)}{s-q^{2}}+\mbox{subtractions},
\end{eqnarray}
where $M_{H}$ is the mass of the hadronic resonance, and
$\lambda_{H}$ gives the coupling of the current to the hadron
$\langle0|j|H\rangle=\lambda_{H}$. In the operator product expansion (OPE) side, the correlator
can be written as
\begin{eqnarray}
\Pi(q^{2})=\int_{s_{min}}^{\infty}ds\frac{\rho^{\mbox{OPE}}(s)}{s-q^{2}}+\Pi^{\mbox{cond}}(q^{2}),
\end{eqnarray}
where the spectral density is $\rho^{\mbox{OPE}}(s)=\frac{1}{\pi}\mbox{Im}\Pi^{\mbox{OPE}}(s)$, with
the integration limit
$s_{min}\approx0$ for $\omega\omega$ state, $s_{min}=(2m_{s})^{2}$ for $\omega\phi$ state, and
$s_{min}=(4m_{s})^{2}$ for $\phi\phi$ state.
After equating the two sides, assuming quark-hadron duality, and
making a Borel transform, the sum rule can be written as
\begin{eqnarray}\label{sr1}
\lambda_{H}^{2}e^{-M_{H}^{2}/M^{2}}&=&\int_{s_{min}}^{s_{0}}ds\rho^{\mbox{OPE}}e^{-s/M^{2}}+\hat{B}\Pi^{\mbox{cond}},
\end{eqnarray}
where $M^2$ indicates the Borel parameter.
To eliminate the hadronic
coupling constant $\lambda_{H}$, one reckons the ratio of derivative
of the sum rule to itself, and then yields
\begin{eqnarray}\label{sum rule 1}
M_{H}^{2}&=&\Bigg\{\int_{s_{min}}^{s_{0}}ds\rho^{\mbox{OPE}}s
e^{-\frac{s}{M^{2}}}+\frac{d(\hat{B}\Pi^{\mbox{cond}})}{d(-\frac{1}{M^{2}})}\Bigg\}\Bigg/\Bigg\{
\int_{s_{min}}^{s_{0}}ds\rho^{\mbox{OPE}}e^{-\frac{s}{M^{2}}}+\hat{B}\Pi^{\mbox{cond}}\Bigg\}.
\end{eqnarray}

For the OPE calculations, we work at the leading order in $\alpha_{s}$ and
consider condensates up to dimension ten, utilizing the light-quark
propagator in the coordinate-space
\begin{eqnarray}
S_{ab}(x)&=&\frac{i\delta_{ab}}{2\pi^{2}x^{4}}\rlap/x-\frac{m_{q}\delta_{ab}}{4\pi^{2}x^{2}}-\frac{i}{32\pi^{2}x^{2}}t^{A}_{ab}gG^{A}_{\mu\nu}(\rlap/x\sigma^{\mu\nu}
+\sigma^{\mu\nu}\rlap/x)-\frac{\delta_{ab}}{12}\langle\bar{q}q\rangle+\frac{i\delta_{ab}}{48}m_{q}\langle\bar{q}q\rangle\rlap/x\nonumber\\&&{}\hspace{-0.3cm}
-\frac{x^{2}\delta_{ab}}{3\cdot2^{6}}\langle g\bar{q}\sigma\cdot Gq\rangle
+\frac{ix^{2}\delta_{ab}}{2^{7}\cdot3^{2}}m_{q}\langle g\bar{q}\sigma\cdot Gq\rangle\rlap/x-\frac{x^{4}\delta_{ab}}{2^{10}\cdot3^{3}}\langle\bar{q}q\rangle\langle g^{2}G^{2}\rangle.\nonumber
\end{eqnarray}
The $s$ quark is
dealt as a light one and the diagrams are considered up to
the order $m_{s}$. For some minor multi-gluon condensate contributions,
one could omit them as the usual treatment.
Concretely, spectral densities can be written as
\begin{eqnarray}
\rho^{\mbox{pert}}(s)=\frac{1}{5\cdot2^{12}\pi^{6}}s^{4},~~~
\rho^{\langle\bar{q}q\rangle^{2}}(s)=\frac{\langle\bar{q}q\rangle^{2}}{2^{3}\pi^{2}}s,~~~
\rho^{\langle\bar{q}q\rangle\langle g\bar{q}\sigma\cdot G q\rangle}(s)=-\frac{\langle\bar{q}q\rangle\langle g\bar{q}\sigma\cdot G q\rangle}{2^{3}\pi^{2}},\nonumber
\end{eqnarray}
\begin{eqnarray}
\hat{B}\Pi^{\mbox{cond}}=\frac{\langle g\bar{q}\sigma\cdot G q\rangle\langle g\bar{q}\sigma\cdot G q\rangle}{2^{6}\pi^{2}}+\frac{\langle\bar{q}q\rangle^{2}\langle g^{2}G^{2}\rangle}{3^{2}\cdot2^{5}\pi^{2}},\nonumber
\end{eqnarray}
for $\omega\omega$ state,
\begin{eqnarray}
\rho^{\mbox{pert}}(s)=\frac{1}{5\cdot2^{12}\pi^{6}}s^{4},~~~
\rho^{\langle\bar{s}s\rangle^{2}}(s)=\frac{\langle\bar{s}s\rangle^{2}}{2^{4}\pi^{2}}s,~~~
\rho^{\langle\bar{q}q\rangle^{2}}(s)=\frac{\langle\bar{q}q\rangle^{2}}{2^{4}\pi^{2}}s,~~~
\rho^{\langle g\bar{s}\sigma\cdot G s\rangle}(s)=\frac{\langle
g\bar{s}\sigma\cdot G
s\rangle}{2^{7}\pi^{4}}m_{s}s,\nonumber
\end{eqnarray}
\begin{eqnarray}
\rho^{\langle\bar{s}s\rangle\langle g^{2}G^{2}\rangle}(s)=-\frac{\langle\bar{s}s\rangle\langle
g^{2}G^{2}\rangle}{3\cdot2^{8}\pi^{4}}m_{s},~~~
\rho^{\langle\bar{s}s\rangle\langle g\bar{s}\sigma\cdot G s\rangle}(s)=-\frac{\langle\bar{s}s\rangle\langle g\bar{s}\sigma\cdot G s\rangle}{2^{4}\pi^{2}},~~~
\rho^{\langle\bar{q}q\rangle\langle g\bar{q}\sigma\cdot G q\rangle}(s)=-\frac{\langle\bar{q}q\rangle\langle g\bar{q}\sigma\cdot G q\rangle}{2^{4}\pi^{2}}\nonumber
\end{eqnarray}
\begin{eqnarray}
\hat{B}\Pi^{\mbox{cond}}=-\frac{m_{s}}{2}\langle\bar{q}q\rangle^{2}\langle\bar{s}s\rangle+\frac{\langle g\bar{s}\sigma\cdot G s\rangle^{2}}{2^{7}\pi^{2}}+\frac{\langle g\bar{q}\sigma\cdot G q\rangle^{2}}{2^{7}\pi^{2}}+\frac{\langle\bar{s}s\rangle^{2}\langle g^{2}G^{2}\rangle}{3^{2}\cdot2^{6}\pi^{2}}+\frac{\langle\bar{q}q\rangle^{2}\langle g^{2}G^{2}\rangle}{3^{2}\cdot2^{6}\pi^{2}},\nonumber
\end{eqnarray}
for $\omega\phi$ state, and
\begin{eqnarray}
\rho^{\mbox{pert}}(s)=\frac{1}{5\cdot2^{12}\pi^{6}}s^{4},~~~
\rho^{\langle\bar{s}s\rangle^{2}}(s)=\frac{\langle\bar{s}s\rangle^{2}}{2^{3}\pi^{2}}s,~~~
\rho^{\langle g\bar{s}\sigma\cdot G s\rangle}(s)=\frac{\langle
g\bar{s}\sigma\cdot G
s\rangle}{2^{6}\pi^{4}}m_{s}s,\nonumber
\end{eqnarray}
\begin{eqnarray}
\rho^{\langle\bar{s}s\rangle\langle g^{2}G^{2}\rangle}(s)=-\frac{\langle\bar{s}s\rangle\langle
g^{2}G^{2}\rangle}{3\cdot2^{7}\pi^{4}}m_{s},~~~
\rho^{\langle\bar{s}s\rangle\langle g\bar{s}\sigma\cdot G s\rangle}(s)=-\frac{\langle\bar{s}s\rangle\langle g\bar{s}\sigma\cdot G s\rangle}{2^{3}\pi^{2}},\nonumber
\end{eqnarray}
\begin{eqnarray}
\hat{B}\Pi^{\mbox{cond}}=-m_{s}\langle\bar{s}s\rangle^{3}+\frac{\langle g\bar{s}\sigma\cdot G s\rangle\langle g\bar{s}\sigma\cdot G s\rangle}{2^{6}\pi^{2}}+\frac{\langle\bar{s}s\rangle^{2}\langle g^{2}G^{2}\rangle}{3^{2}\cdot2^{5}\pi^{2}},\nonumber
\end{eqnarray}
for $\phi\phi$ state.

\section{Numerical analysis and discussions}\label{sec3}
The sum rule (\ref{sum rule 1}) is numerically analyzed in this section. The input values are taken as
$m_{s}=0.10^{+0.03}_{-0.02}~\mbox{GeV}$ \cite{PDG},
$\langle\bar{q}q\rangle=-(0.23\pm0.03)^{3}~\mbox{GeV}^{3}$, $\langle
g\bar{q}\sigma\cdot G q\rangle=m_{0}^{2}~\langle\bar{q}q\rangle$,
$\langle\bar{s}s\rangle=-(0.8\pm0.1)\times(0.23\pm0.03)^{3}~\mbox{GeV}^{3}$, $\langle
g\bar{s}\sigma\cdot G s\rangle=m_{0}^{2}~\langle\bar{s}s\rangle$,
$m_{0}^{2}=0.8\pm0.1~\mbox{GeV}^{2}$, and $\langle
g^{2}G^{2}\rangle=0.88~\mbox{GeV}^{4}$
\cite{overview2}.
Complying with the criterion
of sum rule analysis, the threshold $\sqrt{s_{0}}$ and Borel
parameter $M^{2}$ are varied to find the optimal stability window.
In the standard QCD sum rule approach, one can analyse the convergence in the
OPE side and the pole contribution dominance in the phenomenological
side to determine the conventional Borel window: on one hand, the lower
constraint for $M^{2}$ is obtained by the consideration that the
perturbative contribution should be larger than condensate
contributions; on the other
hand, the upper bound for $M^{2}$ is obtained by the restriction
that the pole contribution should be larger than the continuum
state contributions.
Meanwhile, the threshold
$\sqrt{s_{0}}$ is not arbitrary but characterizes the
beginning of continuum states.
For many hadrons,
the first excitation of studied state defines the size of $\sqrt{s_0}$,
and the difference between $\sqrt{s_0}$ and the mass $M_{H}$ of studied state
is around $0.5~\mbox{GeV}$.
Concretely, the value of $\sqrt{s_{0}}$ is fixed by these steps:
1) taking a value of $\sqrt{s_{0}}$;
2) fixing the corresponding Borel parameters $M^{2}$ according to two rules
(OPE convergence and pole dominance);
3) extracting the mass from the sum
rule in the work window fixed in the first and second steps;
4) checking that whether the $\sqrt{s_{0}}$ chosen in the first step is acceptable using
the empirical relation that the difference between $\sqrt{s_0}$ and $M_{H}$
is around $0.5~\mbox{GeV}$;
5) if the $\sqrt{s_{0}}$ chosen in the first step is not acceptable, return to the first step,
vary $\sqrt{s_{0}}$ and go on.
Taking $\omega\omega$ as an example,
we choose $\sqrt{s_0}=2.4~\mbox{GeV}$ and finally
arrive at $M_{H}=1.97~\mbox{GeV}$. One could check that
$\sqrt{s_0}=2.4~\mbox{GeV}$ is acceptable
with the empirical relation that the difference between $\sqrt{s_0}$ and $M_{H}$
is around $0.5~\mbox{GeV}$.
Thus, we choose the central value of $\sqrt{s_0}=2.4~\mbox{GeV}$ for $\omega\omega$ state.

Here, we would make some particular discussions on the choice of Borel windows.
It has been shown in detail in some Refs, such as
\cite{qqqq} and \cite{Matheus}, that it is not possible to find an conventional Borel window for some
light scalar tetraquarks. The problem is that the four-quark condensate is very large,
making the standard OPE convergence to happen only at very large values of $M^2$.
In fact, it has appeared the same problem in this work.
Taking the $\omega\omega$ as an example,
the comparison
between pole and continuum contributions from sum rule (\ref{sr1}) for $\omega\omega$ state
for $\sqrt{s_{0}}=2.4~\mbox{GeV}$ is shown in the left panel of FIG. 1, and its OPE convergence by comparing the
perturbative with other condensate contributions is shown in the right panel.
Even if we choose some uncritical convergence criteria, e.g. the perturbative contribution should be at least
bigger than each condensate contribution,
there is no standard OPE convergence up to $M^{2}\geq1.8~\mbox{GeV}^{2}$.
The consequence is that it is unable to find the conventional Borel window
where both the OPE converges well (i.e. the perturbative contribution
bigger than each condensate contribution) and the pole dominates over the continuum
(the latter one happens at $M^{2}\leq1.3~\mbox{GeV}^{2}$).
Under such a circumstance, one could try several possible ways to solve the problem.
I) One could release the criterion of pole dominating over continuum and
take some high values of Borel parameter $M^{2}$.
Thus, OPE series can converge well. Graphically from the Borel curve, one can see that there is a very stable plateau.
Some authors have virtually adopted this way to deal with the above problem.
However, there occurs some other problem.
Although there is very good OPE convergence and a flat plateau for the Borel curve,
contributions from continuum states are dominating.
As one knows, the phenomenological side of the sum rule can be expressed as
$\Pi(q^{2})=\frac{\lambda_{H}^{2}}{M_{H}^{2}-q^{2}}+\frac{1}{\pi}\int_{s_{0}}
^{\infty}ds\frac{\mbox{Im}\Pi^{\mbox{phen}}(s)}{s-q^{2}}$
due to the ``single-pole+continuum states" hypothesis.
From the criterion of pole dominating over continuum, one can obtain the maximal value of the Borel parameter $M^{2}$
satisfying the ``single-pole+continuum states" model.
Exceeding this value of $M^{2}$, the single-pole dominance condition will be spoiled.
Thereby, the Borel parameter $M^{2}$ must not be chosen too large to warrant pole dominance.
II) One could push the threshold
parameter $\sqrt{s_{0}}$ to a very large value, and the maximum value of $M^{2}$ will
be enhanced with the increasing of $\sqrt{s_{0}}$.
Thus, one may find the Borel window satisfying both the
perturbative bigger than condensate
contributions and the pole bigger than continuum
contributions.
However, the threshold parameter
$\sqrt{s_{0}}$ is not arbitrary but characterizes the
beginning of the continuum states.
With too large values of $\sqrt{s_{0}}$,
contributions from high resonance states and continuum states may be included in
the pole contribution.
Hence, the QCD sum rule may not work normally.
III) One could warrant the pole dominance firstly and try
releasing the strict convergence criterion of perturbative contribution
larger than each condensate contribution in some case.
In the present work, we have dealt with the problem in this way.
It is worth to note that the treatment is not arbitrary but there
is some definite condition.
For example, we consider the ratio of perturbative contribution
to the ``total OPE contribution" (the sum of perturbative and other condensate
contributions calculated) but not the ratio of perturbative contribution to each condensate contribution.
Not too bad, there are two main condensate contributions with different signs (four-quark condensate
and two-quark multiply mixed
condensate) and they could cancel with each other to some extent, which
brings that the ratio of perturbative contribution
to the ``total OPE contribution" is bigger than $60\%$ at $M^{2}\geq0.8~\mbox{GeV}^{2}$ for $\omega\omega$
for $\sqrt{s_{0}}=2.4~\mbox{GeV}$.
In addition, we calculate and find that the ratio
of perturbative contribution to the ``total OPE contribution" does not
change much including some high dimension condensate contributions.
In this sense,
we could say that the OPE converges in the region while satisfying pole dominance.
Although it may not be good OPE convergence in comparison with the conventional
case, one could find a comparatively reasonable work window.
Note that the treatment should not be arbitrarily transplanted
to any case.
One could take the Ref. \cite{Matheus} as an example.
From its FIG. 4, one can see that for $f_{0}$ all condensate contributions
in the region $0.4\leq M^{2}\leq0.7~\mbox{GeV}^{2}$
are larger than the perturbative contribution.
That means one could not even find a region that the perturbative dominates in
the ``total OPE" allowed by the upper bound.
In a word, to deal with the problem on choosing the conventional Borel window in
QCD sum rules, which have similarly appeared in some other multiquark states, we
warrant the pole dominance preferably and release the strict convergence criterion
to a weak one that perturbative dominates
in ``total OPE contribution", so that the convergence of OPE is still under control while satisfying
pole dominance.
Although it may not be so good OPE convergence as the conventional
case, one could find a comparatively reasonable work window
and extract the hadronic information of studied states reliably.
Thus,
we choose the minimum value of $M^{2}$ to be $0.8~\mbox{GeV}^{2}$
and the maximum $M^{2}$ to be
$1.3~\mbox{GeV}^{2}$ for $\omega\omega$ state for $\sqrt{s_0}=2.4~\mbox{GeV}$. Similarly,
the maximum value of $M^{2}$ is
taken as $1.2~\mbox{GeV}^{2}$ for $\sqrt{s_0}=2.3~\mbox{GeV}$;
for $\sqrt{s_0}=2.5~\mbox{GeV}$, the maximum $M^{2}$ is taken as $1.4~\mbox{GeV}^{2}$.
The
dependence on $M^2$ for the mass of $\omega\omega$ state from sum rule (\ref{sum rule 1}) is
shown in FIG. 2, and we arrive at
$1.97\pm0.07~\mbox{GeV}$ for $\omega\omega$ state. Considering the uncertainty rooting in the variation of quark masses and
condensates, we gain
$1.97\pm0.07\pm0.10~\mbox{GeV}$ (the
first error reflects the uncertainty due to variation of $\sqrt{s_{0}}$
and $M^{2}$, and the second error resulted from the variation of
QCD parameters) or $1.97\pm0.17~\mbox{GeV}$
for $\omega\omega$ state.

The comparison
between pole and continuum contributions from sum rule (\ref{sr1}) for $\omega\phi$ state
for $\sqrt{s_{0}}=2.6~\mbox{GeV}$ is shown in the left panel of FIG. 3, and its OPE convergence by comparing the
perturbative with other condensate contributions is shown in the right panel.
There has the same problem for $\omega\phi$ as the above case for $\omega\omega$, and we treat
it similarly. For $\omega\phi$ state, the ratio of
perturbative  to the ``total OPE contribution" at $M^{2}=1.1~\mbox{GeV}^{2}$ for $\sqrt{s_{0}}=2.6~\mbox{GeV}$ is around $67\%$
and increases with the $M^{2}$. Furthermore, the relative pole contribution
is approximate to $54\%$ at $M^{2}=1.5~\mbox{GeV}^{2}$ and descends with the $M^{2}$.
Thus, the range of $M^{2}$ is taken as
$M^{2}=1.1\sim1.5~\mbox{GeV}^{2}$ for $\sqrt{s_0}=2.6~\mbox{GeV}$.
Similarly,
the proper range of $M^{2}$ is obtained as $1.1\sim1.4~\mbox{GeV}^{2}$ for $\sqrt{s_0}=2.5~\mbox{GeV}$, and
the range of $M^{2}$ is $1.1\sim1.6~\mbox{GeV}^{2}$ for $\sqrt{s_0}=2.7~\mbox{GeV}$.
The mass of $\omega\phi$ state as a function of $M^{2}$ from sum rule
(\ref{sum rule 1}) is
shown in FIG. 4, and we obtain
$2.07\pm0.13~\mbox{GeV}$ for $\omega\phi$. Varying input values of quark masses and
condensates, we attain
$2.07\pm0.13\pm0.08~\mbox{GeV}$ (the
first error reflects the uncertainty due to variation of $\sqrt{s_{0}}$
and $M^{2}$, and the second error resulted from the variation of
QCD parameters) or $2.07\pm0.21~\mbox{GeV}$
for $\omega\phi$ state.

For $\phi\phi$ state, the comparison
between pole and continuum contributions from sum rule (\ref{sr1})
for $\sqrt{s_{0}}=2.7~\mbox{GeV}$ is shown as an example in the left panel of FIG. 5, and its OPE convergence by comparing the
perturbative with other condensate contributions is shown in the right panel.
A bit difference for the case of $\phi\phi$ is that
the perturbative contribution can be
bigger than the second most important condensate $\langle \bar{s}s\rangle^{2}$ at $M^{2}\geq1.1~\mbox{GeV}^{2}$.
Meanwhile, the pole contribution can dominate in the total contribution while $M^{2}\leq1.6~\mbox{GeV}^{2}$.
Thus, it is possible to find a region
where both the OPE can converge well (the perturbative contribution
bigger than each condensate contribution) and the pole dominates over the continuum.
Thus, the range of $M^{2}$ for $\phi\phi$ state is taken as
$M^{2}=1.1\sim1.6~\mbox{GeV}^{2}$ for $\sqrt{s_0}=2.7~\mbox{GeV}$.
Via the similar analyzing process,
the proper range of $M^{2}$ is obtained as $1.1\sim1.7~\mbox{GeV}^{2}$ for $\sqrt{s_0}=2.8~\mbox{GeV}$, and
the range of $M^{2}$ is $1.1\sim1.8~\mbox{GeV}^{2}$ for $\sqrt{s_0}=2.9~\mbox{GeV}$.
In the chosen region,
the corresponding Borel curve to determine the mass of $\phi\phi$ state is shown in FIG. 6, and
we extract the mass value $2.18\pm0.20~\mbox{GeV}$
for $\phi\phi$ state. Subsequently,
we vary the quark masses as well as
condensates and arrive at $2.18\pm0.20\pm0.09~\mbox{GeV}$ (the
first error reflects the uncertainty due to variation of $\sqrt{s_{0}}$
and $M^{2}$, and the second error resulted from the variation of
QCD parameters) or $2.18\pm0.29~\mbox{GeV}$ in a concise form.

\begin{figure}
\centerline{\epsfysize=5.0truecm\epsfbox{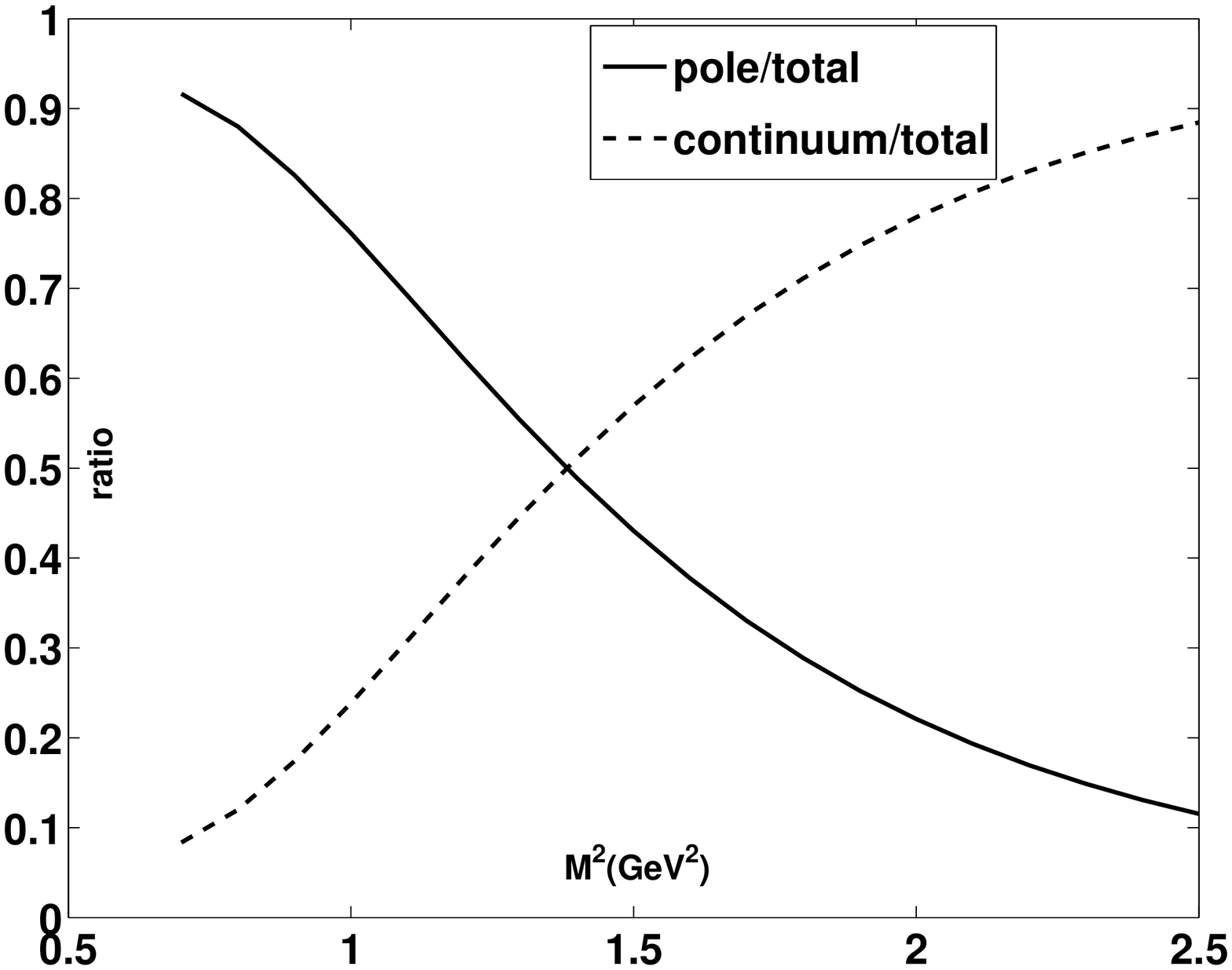}\epsfysize=5.0truecm\epsfbox{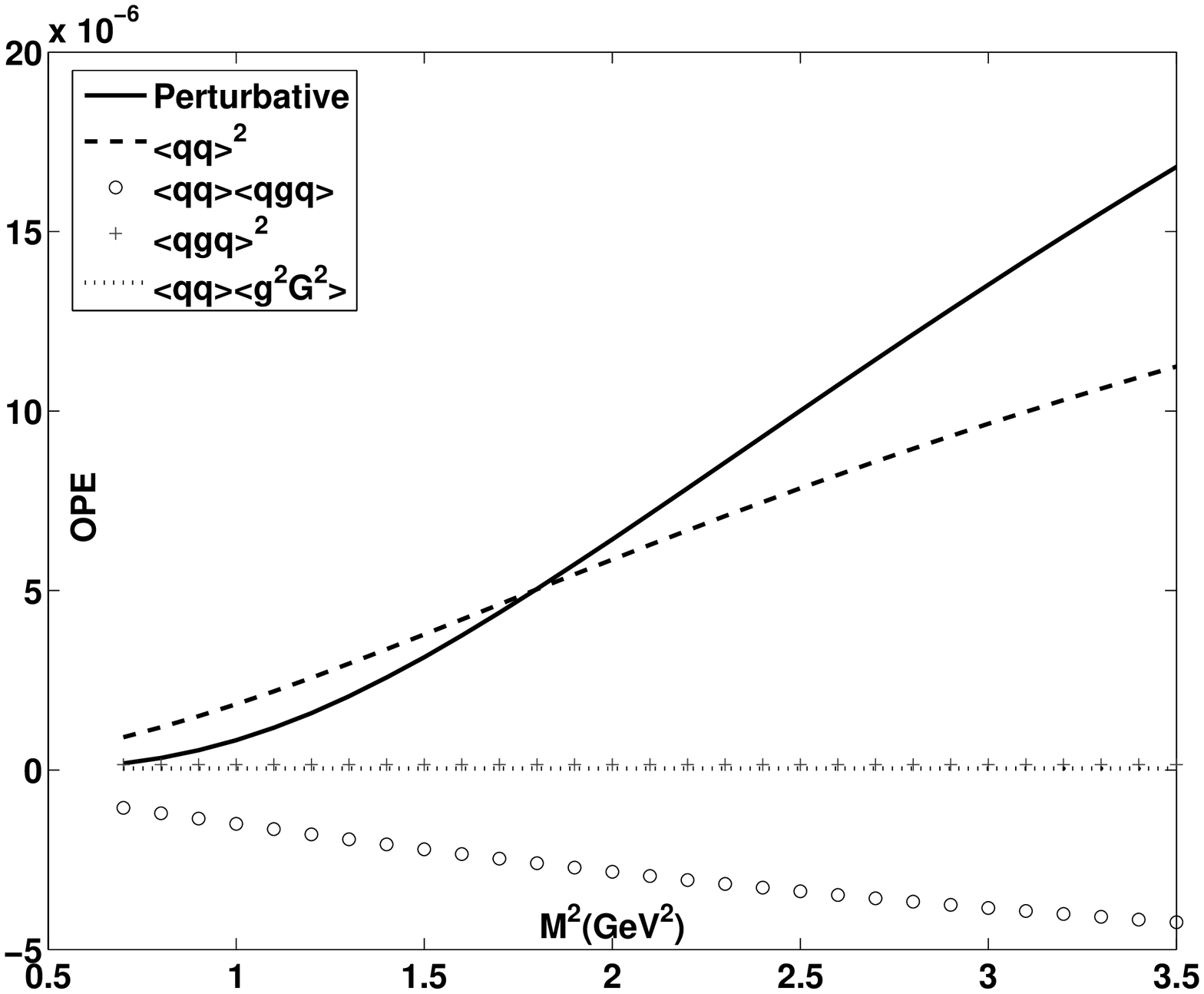}}
\caption{In the left panel, the solid line shows the relative pole contribution
(the pole contribution divided by the total, pole plus continuum
contribution) and the dashed line shows the relative continuum
contribution from sum rule (\ref{sr1}) for $\sqrt{s_{0}}=2.4~\mbox{GeV}$ for
$\omega\omega$ state. The OPE convergence is shown by comparing the
perturbative with other condensate contributions from sum rule (\ref{sr1}) for $\sqrt{s_{0}}=2.4~\mbox{GeV}$ for
$\omega\omega$ state in the right panel. }
\end{figure}

\begin{figure}
\centerline{\epsfysize=5.0truecm
\epsfbox{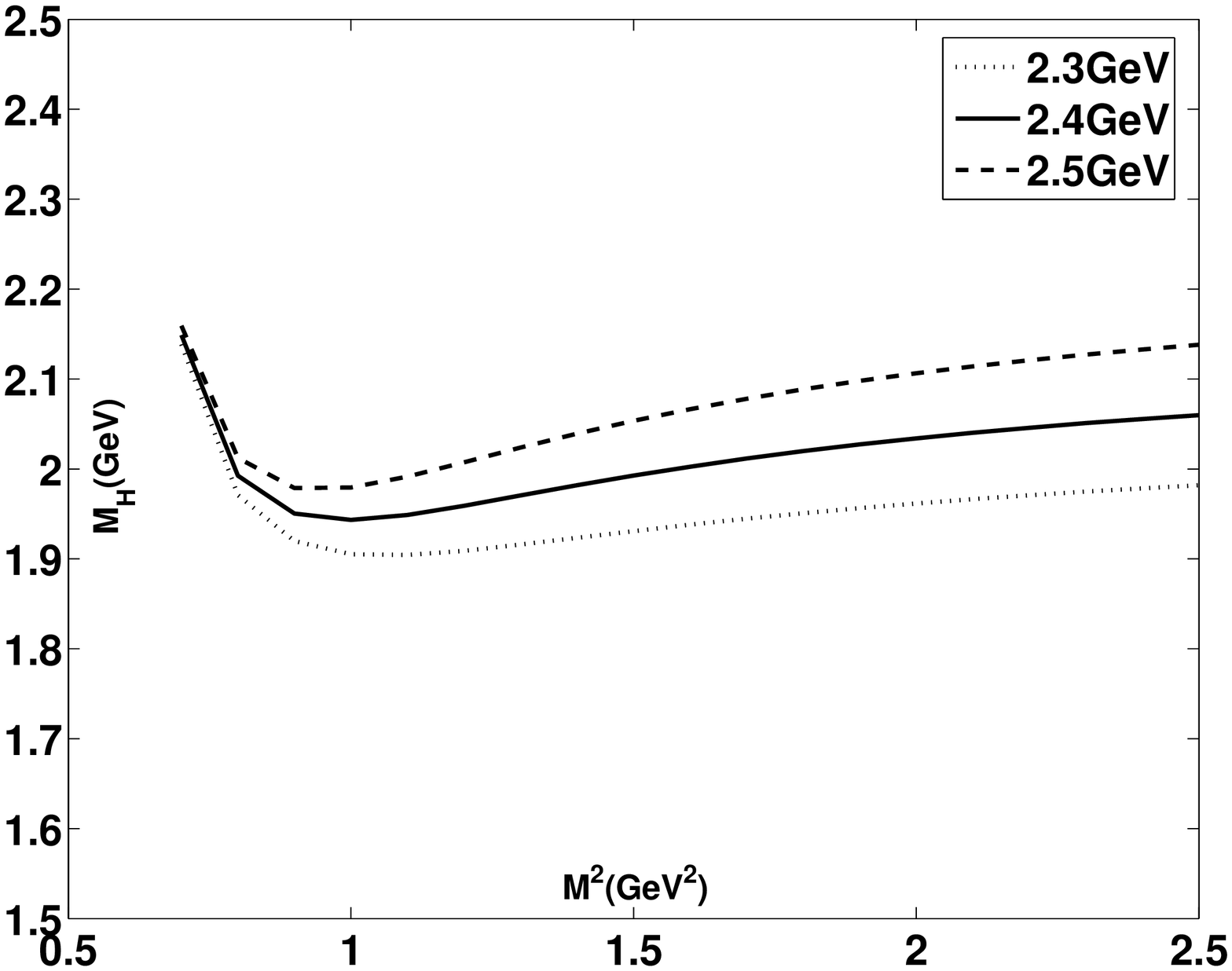}}\caption{The mass of $\omega\omega$ state as a function of $M^{2}$ from sum rule
(\ref{sum rule 1}) is shown. The continuum
thresholds are taken as $\sqrt{s_0}=2.3\sim2.5~\mbox{GeV}$. For
$\sqrt{s_0}=2.3~\mbox{GeV}$, the range of $M^{2}$ is $0.8\sim1.2~\mbox{GeV}^{2}$;
for $\sqrt{s_0}=2.4~\mbox{GeV}$, the range of $M^{2}$ is $0.8\sim1.3~\mbox{GeV}^{2}$;
for $\sqrt{s_0}=2.5~\mbox{GeV}$, the range of $M^{2}$ is $0.8\sim1.4~\mbox{GeV}^{2}$.}
\end{figure}

\begin{figure}
\centerline{\epsfysize=5.0truecm\epsfbox{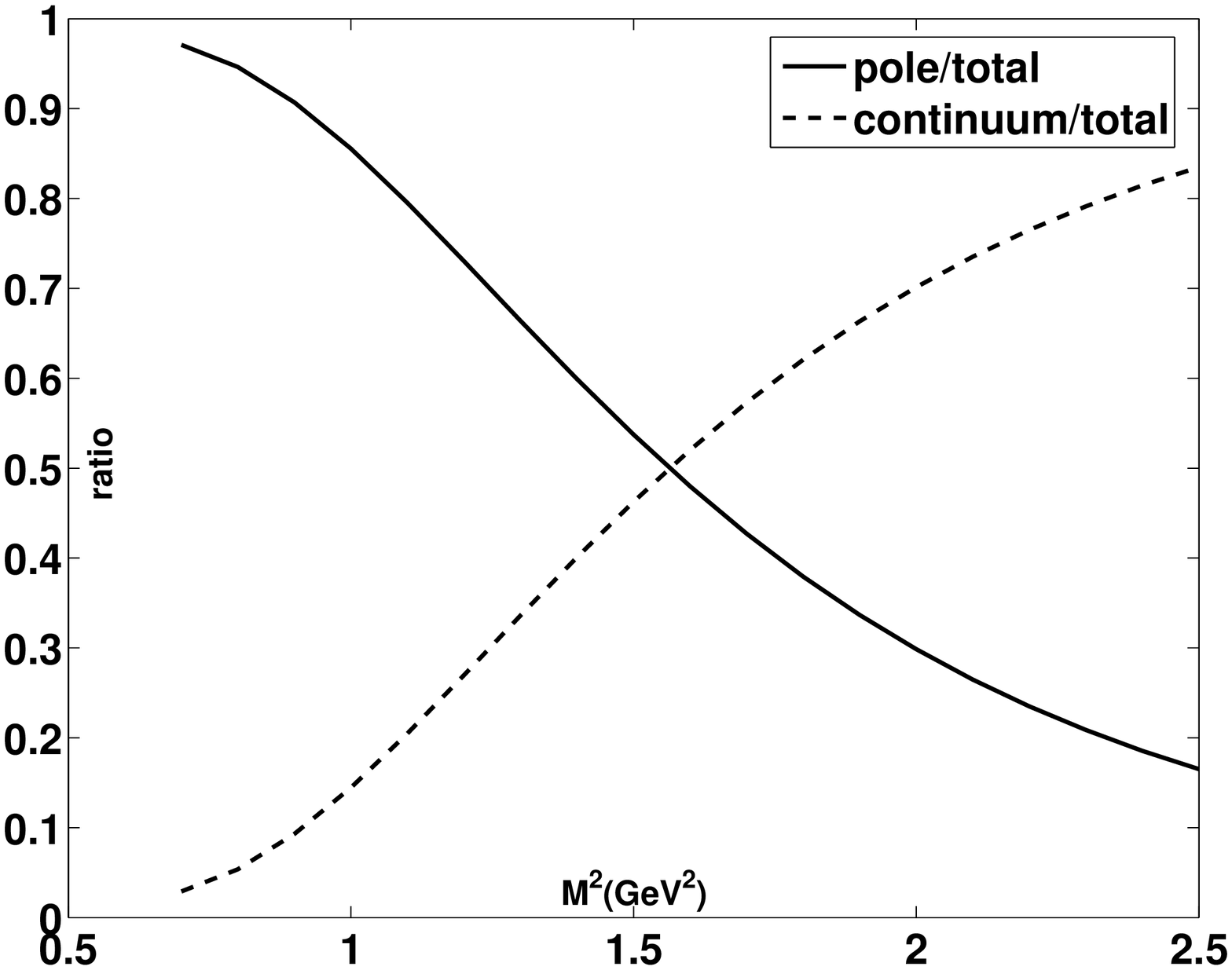}\epsfysize=5.0truecm\epsfbox{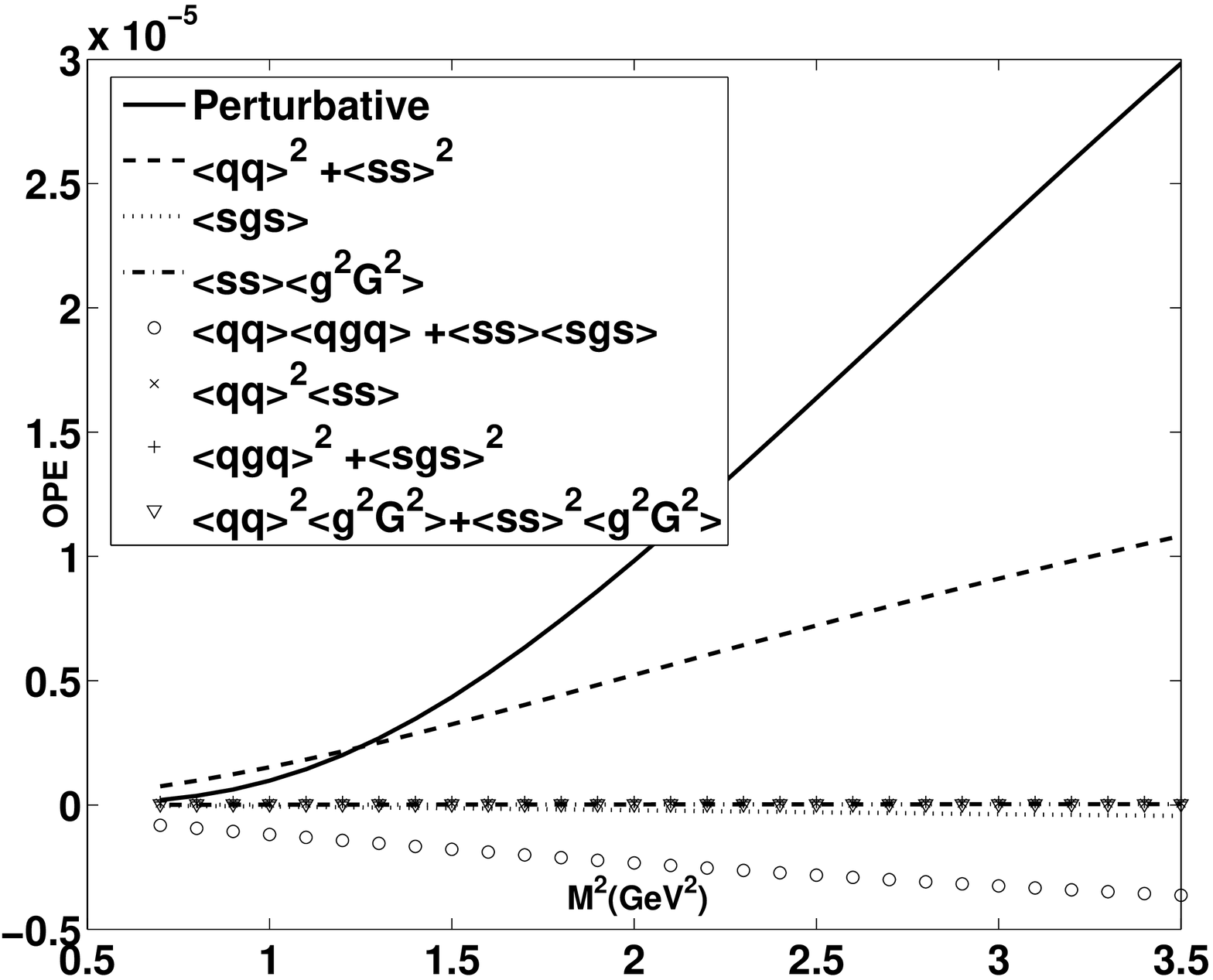}}
\caption{In the left panel, the solid line shows the relative pole contribution
(the pole contribution divided by the total, pole plus continuum
contribution) and the dashed line shows the relative continuum
contribution from sum rule (\ref{sr1}) for $\sqrt{s_{0}}=2.6~\mbox{GeV}$ for
$\omega\phi$ state. The OPE convergence is shown by comparing the
perturbative with other condensate contributions from sum rule (\ref{sr1}) for $\sqrt{s_{0}}=2.6~\mbox{GeV}$ for
$\omega\phi$ state in the right panel. }
\end{figure}

\begin{figure}
\centerline{\epsfysize=5.0truecm
\epsfbox{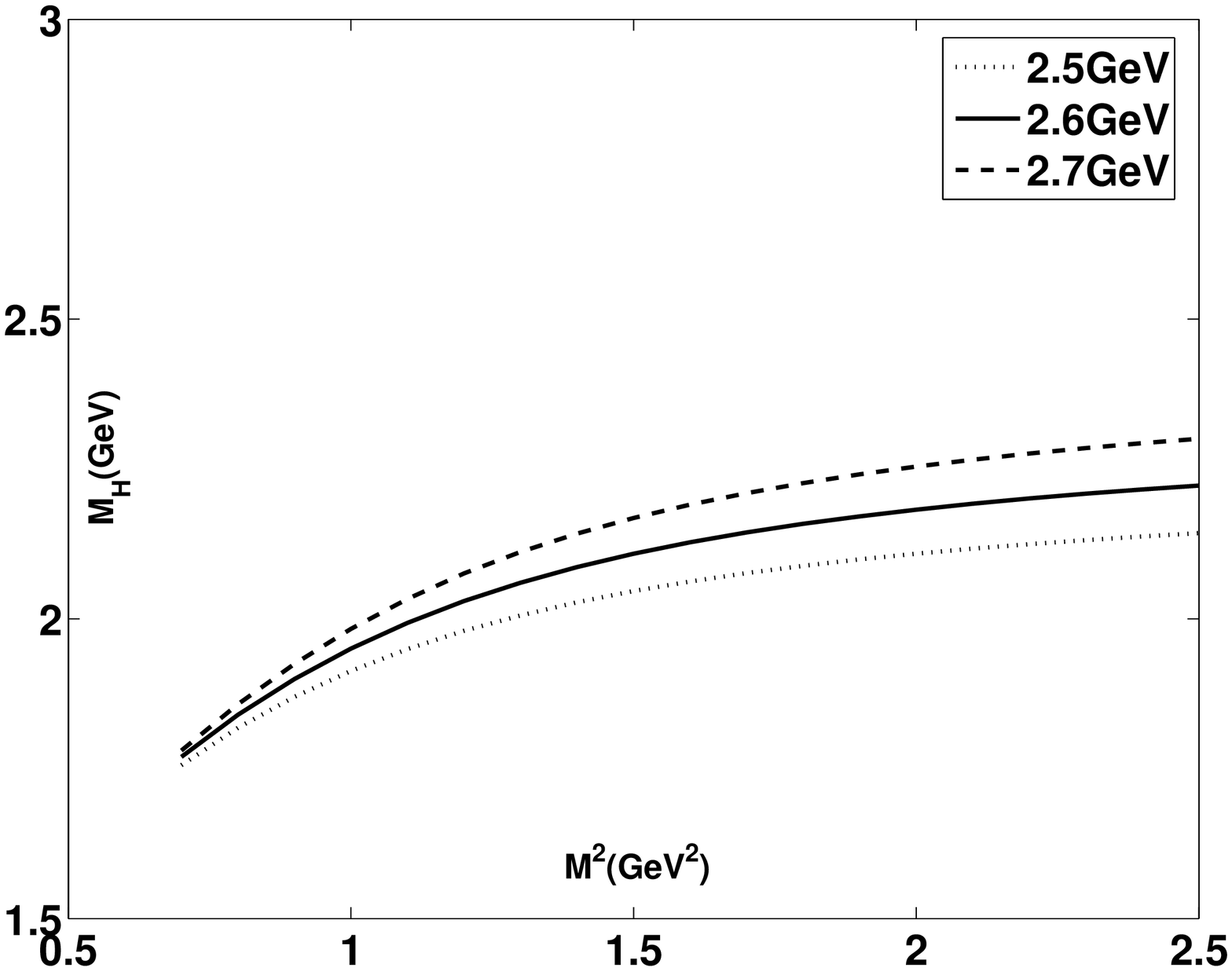}}\caption{The mass of $\omega\phi$ state as a function of $M^{2}$ from sum rule
(\ref{sum rule 1}) is shown in the right panel. The continuum
thresholds are taken as $\sqrt{s_0}=2.5\sim2.7~\mbox{GeV}$. For
$\sqrt{s_0}=2.5~\mbox{GeV}$, the range of $M^{2}$ is $1.1\sim1.4~\mbox{GeV}^{2}$;
for $\sqrt{s_0}=2.6~\mbox{GeV}$, the range of $M^{2}$ is $1.1\sim1.5~\mbox{GeV}^{2}$;
for $\sqrt{s_0}=2.7~\mbox{GeV}$, the range of $M^{2}$ is $1.1\sim1.6~\mbox{GeV}^{2}$. }
\end{figure}

\begin{figure}
\centerline{\epsfysize=5.0truecm\epsfbox{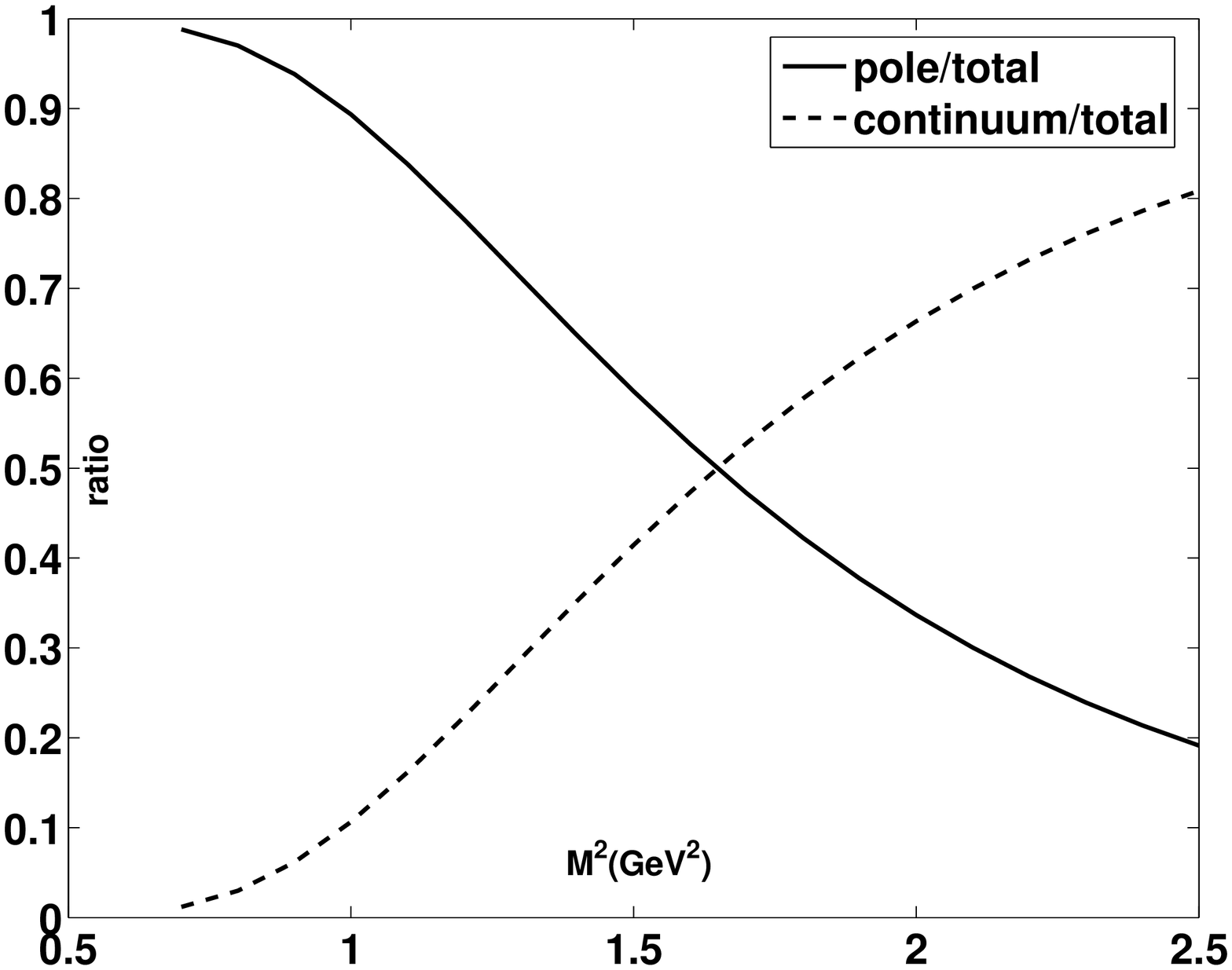}\epsfysize=5.0truecm\epsfbox{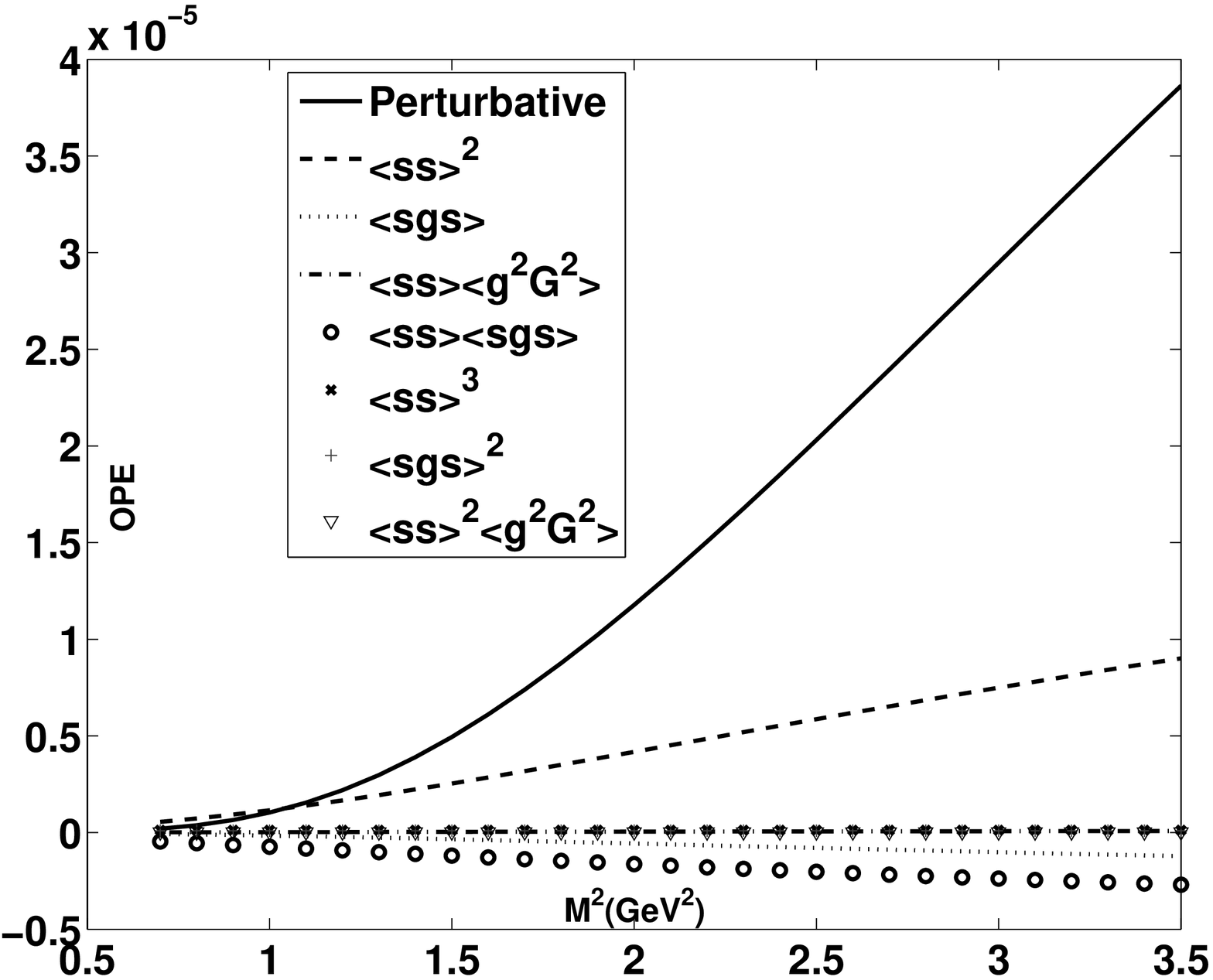}}
\caption{In the left panel, the solid line shows the relative pole contribution
(the pole contribution divided by the total, pole plus continuum
contribution) and the dashed line shows the relative continuum
contribution from sum rule (\ref{sr1}) for $\sqrt{s_{0}}=2.7~\mbox{GeV}$ for
$\phi\phi$ state. The OPE convergence is shown by comparing the
perturbative with other condensate contributions from sum rule (\ref{sr1}) for $\sqrt{s_{0}}=2.7~\mbox{GeV}$ for
$\phi\phi$ state in the right panel. }
\end{figure}

\begin{figure}
\centerline{\epsfysize=5.0truecm
\epsfbox{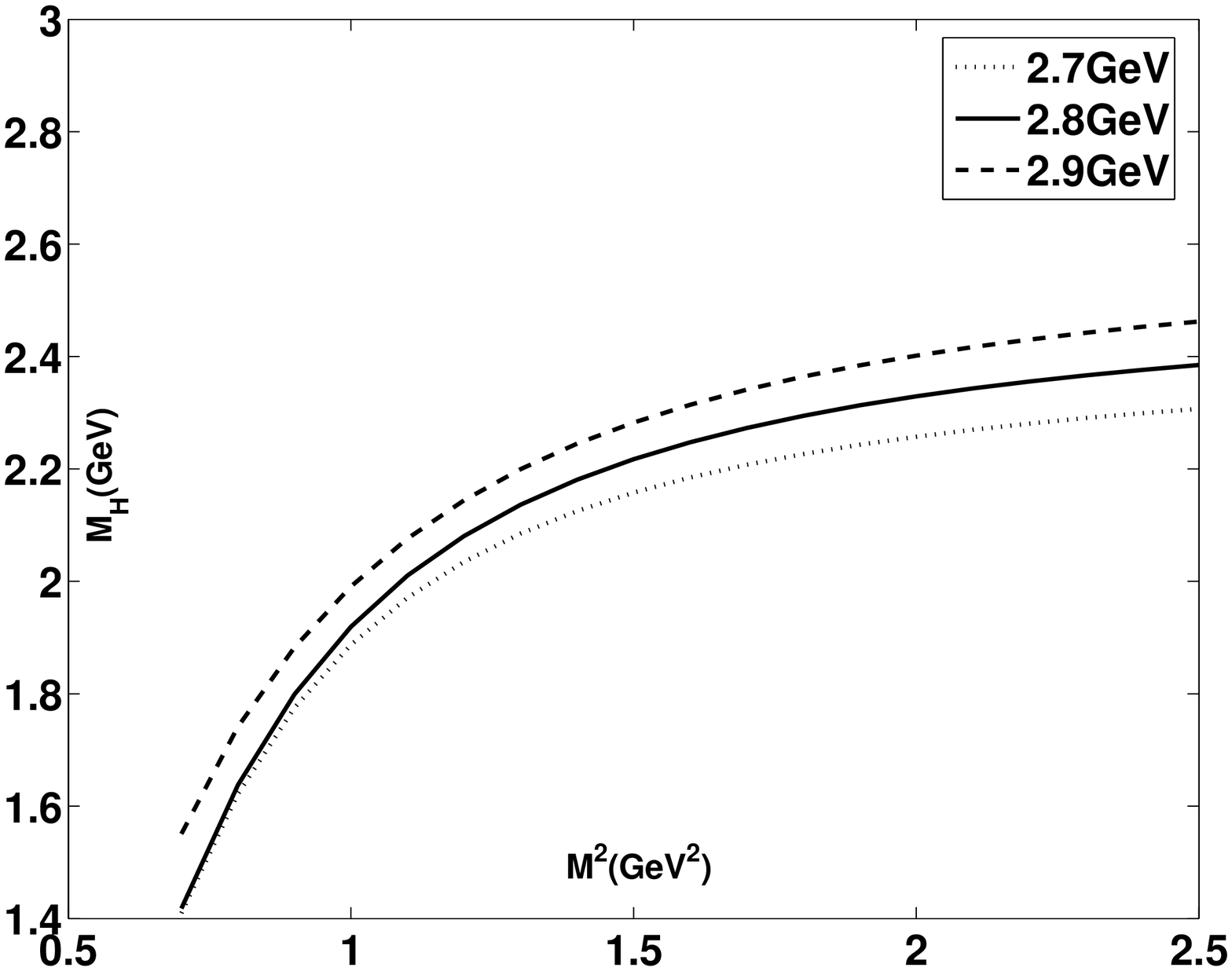}}\caption{The mass of $\phi\phi$ state as a function of $M^{2}$ from sum rule
(\ref{sum rule 1}) is shown. The continuum
thresholds are taken as $\sqrt{s_0}=2.7\sim2.9~\mbox{GeV}$. For
$\sqrt{s_0}=2.7~\mbox{GeV}$, the range of $M^{2}$ is $1.1\sim1.6~\mbox{GeV}^{2}$;
for $\sqrt{s_0}=2.8~\mbox{GeV}$, the range of $M^{2}$ is $1.1\sim1.7~\mbox{GeV}^{2}$;
for $\sqrt{s_0}=2.9~\mbox{GeV}$, the range of $M^{2}$ is $1.1\sim1.8~\mbox{GeV}^{2}$. }
\end{figure}

\section{Summary and outlook}\label{sec4}
In $\gamma\gamma\rightarrow X\rightarrow\omega\omega$,
$\omega\phi$,
and $\phi\phi$,
Belle Collaboration observed three new resonant structures at
$M(\omega\omega)\sim1.91~\mbox{GeV}$,
$M(\omega\phi)\sim2.2~\mbox{GeV}$,
and $M(\phi\phi)\sim2.35~\mbox{GeV}$.
Assuming these newly observed resonances as
molecular states, we have employed the QCD sum rule method to calculate their
masses, taking into account contributions of operators up to dimension ten in the OPE.
Our final numerical
results are $1.97\pm0.17~\mbox{GeV}$ for $\omega\omega$ state, $2.07\pm0.21~\mbox{GeV}$
for $\omega\phi$ state, and
$2.18\pm0.29~\mbox{GeV}$
for $\phi\phi$ state,
which are in agreement with the experimental values of $X(1910)$, $X(2200)$, and $X(2350)$ respectively.
This supports the statement that
$X(1910)$, $X(2200)$, and $X(2350)$ could be
$\omega\omega$, $\omega\phi$, and $\phi\phi$ molecular states respectively.
However, one should note that there are still some
differences between our central values and experimental values.
At present, we have merely
considered $\omega\omega$, $\omega\phi$, and $\phi\phi$ molecular states
with $J^{P}=0^{+}$. Belle Collaboration indicated that
while there are substantial spin-$0$
components in all three modes (namely $\gamma\gamma\rightarrow X\rightarrow\omega\omega$,
$\omega\phi$,
and $\phi\phi$), there are also spin-$2$
components near threshold.
The differences between our central values and experimental data are probably caused by that
we have not considered
the spin-$2$ components for $\omega\omega$, $\omega\phi$, and $\phi\phi$ state here,
which implies that the theoretical predictions might be improved
by including $J=2$ components for the future.
In addition, one needs to take into account other
dynamical analysis to identify the nature structures of these $X$ States for further work.

\begin{acknowledgments}
The authors would like to thank the anonymous referees for 
useful suggestions and comments.
The authors also thank H.~X.~Chen for the recent communication and helpful discussions.
This work was supported in part by the National Natural Science
Foundation of China under Contract Nos.11105223, 10947016, and 10975184.
\end{acknowledgments}

\end{document}